\PassOptionsToPackage{unicode}{hyperref}
\PassOptionsToPackage{hyphens}{url}
\PassOptionsToPackage{dvipsnames,svgnames,x11names}{xcolor}
\documentclass[12pt]{article}

\usepackage{amsmath,amssymb, mathtools, bm}
\usepackage{iftex}
\usepackage[edges]{forest}
\usepackage{booktabs}
\usepackage{tabularx}
\usepackage{array}
\usepackage{pifont}
\usepackage{threeparttable}
\usepackage[utf8]{inputenc}   
\usepackage[T1]{fontenc}
\usepackage{mathrsfs}
\usepackage{comment}

\usepackage{booktabs}
\usepackage{tabularx}
\usepackage{longtable}
\usepackage{threeparttablex} 
\usepackage{pifont} 

\usepackage{array}
\usepackage{makecell}
\usepackage{xltabular} %
\usepackage{ragged2e}  
\usepackage{multirow}
\usepackage{graphicx}

\ifPDFTeX
  \usepackage[T1]{fontenc}
  \usepackage[utf8]{inputenc}
  \usepackage{textcomp} 
\else 
  \usepackage{unicode-math}
  \defaultfontfeatures{Scale=MatchLowercase}
  \defaultfontfeatures[\rmfamily]{Ligatures=TeX,Scale=1}
\fi
\usepackage{lmodern}
\ifPDFTeX\else  
\fi
\IfFileExists{upquote.sty}{\usepackage{upquote}}{}
\IfFileExists{microtype.sty}{
  \usepackage[]{microtype}
  \UseMicrotypeSet[protrusion]{basicmath} 
}{}
\makeatletter
\@ifundefined{KOMAClassName}{
  \IfFileExists{parskip.sty}{%
    \usepackage{parskip}
  }{
    \setlength{\parindent}{0pt}
    \setlength{\parskip}{6pt plus 2pt minus 1pt}}
}{
  \KOMAoptions{parskip=half}}
\makeatother
\usepackage{xcolor}
\makeatletter
\ifx\paragraph\undefined\else
  \let\oldparagraph\paragraph
  \renewcommand{\paragraph}{
    \@ifstar
      \xxxParagraphStar
      \xxxParagraphNoStar
  }
  \newcommand{\xxxParagraphStar}[1]{\oldparagraph*{#1}\mbox{}}
  \newcommand{\xxxParagraphNoStar}[1]{\oldparagraph{#1}\mbox{}}
\fi
\ifx\subparagraph\undefined\else
  \let\oldsubparagraph\subparagraph
  \renewcommand{\subparagraph}{
    \@ifstar
      \xxxSubParagraphStar
      \xxxSubParagraphNoStar
  }
  \newcommand{\xxxSubParagraphStar}[1]{\oldsubparagraph*{#1}\mbox{}}
  \newcommand{\xxxSubParagraphNoStar}[1]{\oldsubparagraph{#1}\mbox{}}
\fi
\makeatother

\usepackage{longtable,booktabs,array}
\usepackage{calc} 
\usepackage{etoolbox}
\makeatletter
\patchcmd\longtable{\par}{\if@noskipsec\mbox{}\fi\par}{}{}
\makeatother
\IfFileExists{footnotehyper.sty}{\usepackage{footnotehyper}}{\usepackage{footnote}}
\makesavenoteenv{longtable}
\usepackage{graphicx}
\makeatletter
\def\maxwidth{\ifdim\Gin@nat@width>\linewidth\linewidth\else\Gin@nat@width\fi}
\def\maxheight{\ifdim\Gin@nat@height>\textheight\textheight\else\Gin@nat@height\fi}
\makeatother

\setkeys{Gin}{width=\maxwidth,height=\maxheight,keepaspectratio}
\makeatletter
\def\fps@figure{htbp}
\makeatother

\makeatletter
\@ifpackageloaded{caption}{}{\usepackage{caption}}
\AtBeginDocument{%
\ifdefined\contentsname
  \renewcommand*\contentsname{Table of contents}
\else
  \newcommand\contentsname{Table of contents}
\fi
\ifdefined\listfigurename
  \renewcommand*\listfigurename{List of Figures}
\else
  \newcommand\listfigurename{List of Figures}
\fi
\ifdefined\listtablename
  \renewcommand*\listtablename{List of Tables}
\else
  \newcommand\listtablename{List of Tables}
\fi
\ifdefined\figurename
  \renewcommand*\figurename{Figure}
\else
  \newcommand\figurename{Figure}
\fi
\ifdefined\tablename
  \renewcommand*\tablename{Table}
\else
  \newcommand\tablename{Table}
\fi
}
\@ifpackageloaded{float}{}{\usepackage{float}}
\floatstyle{ruled}
\@ifundefined{c@chapter}{\newfloat{codelisting}{h}{lop}}{\newfloat{codelisting}{h}{lop}[chapter]}
\floatname{codelisting}{Listing}

\makeatother
\makeatletter
\@ifpackageloaded{caption}{}{\usepackage{caption}}
\@ifpackageloaded{subcaption}{}{\usepackage{subcaption}}
\makeatother

\ifLuaTeX
  \usepackage{selnolig}  
\fi
\usepackage[]{natbib}
\usepackage{bookmark}

\IfFileExists{xurl.sty}{\usepackage{xurl}}{} 
\hypersetup{
  pdftitle={Recent advances in the Bradley--Terry model: theory, algorithms, and applications},
  pdfauthor={Shuxing Fang; Ruijian Han; Yuanhang Luo; Yiming Xu},
  pdfkeywords={algorithms, asymptotic theory, comparison data, preference alignment, ranking, sparsity},
  colorlinks=true,
  linkcolor={blue},
  filecolor={Maroon},
  citecolor={Blue},
  urlcolor={Blue},
  pdfcreator={LaTeX via pandoc}}

\def\A{{\mathbb A}}
\def\P{{\mathbb P}}

\def\R{{\mathbb R}}
\def\HH{{\mathcal H}}
\newcommand{\ER}{\text{Erd\H{o}s--R\'enyi}}
\DeclareMathOperator*{\argmax}{arg\,max}

\newcommand{\YH}[1]{\textcolor{orange}{\small {\sf YH: #1}}}

\newcommand{\lp}{\left(}
\newcommand{\rp}{\right)}

\newcommand{\anon}{1}

\begin{document}

\def\spacingset#1{\renewcommand{\baselinestretch}%
{#1}\small\normalsize} \spacingset{1}


\if1\anon
{
  \title{\bf Recent advances in the Bradley--Terry model: theory, algorithms, and applications}
	\author{Shuxing $\rm{Fang}^{a}$, Ruijian $\rm{Han}^{a}$,
	Yuanhang $\rm{Luo}^{a}$,
    and
	Yiming $\rm{Xu}^{b}$\thanks{Authorships are listed alphabetically. R. Han and Y. Xu are corresponding authors. }\\
	\\
	{\small {\small {$\it^{a}$  Department of Data Science and Artificial Intelligence, The Hong Kong Polytechnic University  } }}\\
	{\small {\small {$\it^{b}$ Department of Mathematics, University of Kentucky} }}
}

  \maketitle
} \fi

\if0\anon
{
  \bigskip
  \bigskip
  \bigskip
  \begin{center}
    {\LARGE\bf Title}
\end{center}
  \medskip
} \fi

\bigskip
\begin{abstract}
This article surveys recent advances in the Bradley--Terry (BT) model and its extensions. We focus on statistical and computational aspects, with particular emphasis on the regime in which both the number of objects and the volume of comparisons tend to infinity—a setting relevant to large-scale applications. The main topics include asymptotic theory for statistical estimation and inference, together with the corresponding computational algorithms. We also review applications of these models in sports analytics, social choice, psychometrics, and machine learning. Finally, we discuss several key challenges and outline directions for future research.
\end{abstract}

\noindent%

\vfill

\newpage
\spacingset{1.0}


\section{Introduction}\label{sec:intro}

The Bradley--Terry (BT) model is a popular parametric model for pairwise comparisons. Introduced by \citet{zermelo1929berechnung} and later formalized by \citet{bradley1952rank}, it has been widely used in fields such as psychometrics, social choice, and sports analytics. Recently, the BT model has attracted growing interest in the machine learning community, particularly in deep reinforcement learning and training large language models.  

Modern applications of the BT model often involve much larger datasets than before. For example, ranking current tennis players along with past legends requires analyzing tens of thousands of players over extended historical periods. In machine learning applications, the number of items being compared can reach hundreds of thousands, often with sparse comparison data. The growth in data dimensionality and sparsity necessitates new statistical theory and computational algorithms for the BT model. These questions have received substantial attention from both the statistics and computer science communities over the past three decades, and significant progress has been made.

Prior surveys and monographs on the BT model primarily focus on foundational methodology \citep{david1988method, agresti1990categorical}, modeling \citep{cattelan2012models}, or historical accounts \citep{hamilton2025many}. To complement these works, we survey recent statistical and computational advances, with a particular emphasis on regimes in which both the number of objects and the number of comparisons grow to infinity. The purpose of this review is to provide a concise and accessible summary of the key technical ingredients for researchers and practitioners from different fields who wish to understand and apply the BT model in their own settings.

Our discussion centers on the BT model for pairwise comparisons. We also consider several important generalizations, including general pairwise comparison models with alternative link functions, the Plackett--Luce (PL) model for multiple comparisons, and mixture and covariate-assisted models that accommodate heterogeneity and covariate information. Our main contribution is to provide a systematic synthesis of recent advances in these models from both theoretical and computational perspectives. To keep the exposition concise and focused, we omit some other important developments that warrant more detailed treatment.

The remainder of the article is organized as follows. Section~\ref{sec:2} provides a brief introduction to the BT model and its extensions. Section~\ref{sec:3} describes the data structure in the multiple-comparison setting and common assumptions. Section~\ref{sec:st} offers a comprehensive overview of statistical methods, including estimation procedures and their asymptotic theory. Section~\ref{sec:alg} reviews existing algorithms, both classical and recent, for parameter estimation. Section~\ref{sec:app} surveys applications of these models and Section~\ref{sec:con} concludes the article. The main notation and relationships among the models are summarized in Table~\ref{tab:notation} and Figure~\ref{fig:model-relations}, respectively, in the Supplementary Material.

\section{BT model and its variants}\label{sec:2}

\subsection{The BT model}

Consider $n$ objects indexed by $[n] \coloneqq \{1, \ldots, n\}$. In the classical pairwise comparison setting, for $\{i, j\} \subseteq [n]$, the comparison outcome $Y_{ij}$ takes values in $\{ i \succ j,\, j \succ i \}$, where $i \succ j$ indicates that object $i$ wins over object $j$. The BT model posits $Y_{ij}$ as a random variable with distribution 
\begin{align}
\P(Y_{ij} = i \succ j )= \frac{\gamma_i}{\gamma_i + \gamma_j}, \tag{BT-S}\label{eq:BT_probs}
\end{align}
where $\gamma_i > 0$ is the \textit{strength}\footnote{This is called \textit{Spielst{\"a}rken} in the original paper by \citet{zermelo1929berechnung} which means ``playing strengths.''} associated with object $i$. The winning probabilities of the objects are thus proportional to their strengths.

Sometimes it is helpful to set $u_i = \log \gamma_i\in\R$, which is called the \textit{utility} of $i$. An advantage of this parametrization is that the positivity constraint is no longer needed. In this case, the winning probability in \eqref{eq:BT_probs} becomes
\begin{equation}
\P(Y_{ij} = i \succ j)= \frac{\exp(u_i)}{\exp(u_i) + \exp(u_j)}= \sigma(u_i - u_j),\tag{BT-U}\label{eq:BT_prob}
\end{equation}
where $\sigma(x) = (1 + \exp(-x))^{-1}$ is the logistic function. We adopt \eqref{eq:BT_prob} as the default parametrization for the BT model unless otherwise specified (e.g., Section~\ref{sec:alg}). The vectors $\bm{\gamma}=(\gamma_1, \ldots, \gamma_n)^\top \in \R_+^n$ and $\bm u = (u_1, \ldots, u_n)^\top \in \R^n$ are the \textit{strength vector} and \emph{utility vector}, respectively. 

Since the likelihood in \eqref{eq:BT_prob} depends only on the difference $u_i - u_j$, the BT model is identifiable only up to an additive constant. Common identifiability constraints include the sum-to-zero condition $\sum_{i\in [n]} u_i = 0$ or the reference condition $u_i = 0$ for a chosen $i \in [n]$ \citep{hunter2004mm,cattelan2012models}. See Section~\ref{sec:ident} for an extended discussion on model identifiability. 

The BT model enjoys close connections to other well-studied statistical models. For instance, it can be viewed as logistic regression (or generalized linear models) by introducing covariate vectors $\bm{x}_{ij} = \bm{e}_i - \bm{e}_j$, where $\bm{e}_i$ denotes the $i$-th canonical basis vector in $\mathbb{R}^n$. This perspective yields a natural generalization of the BT model through the use of alternative link functions, as will be discussed in Section~\ref{sec:gpc}. The BT model can also be derived from the marginal distribution of the PL model for multiple comparisons, which will be discussed in Section~\ref{sec:pl}. Moreover, the BT model maximizes the total pairwise entropy subject to a fixed mean score sequence \citep{joe1988majorization, aldous2022stay}. This observation suggests an analogous role for the BT model to that of the Gaussian distribution and thus establishes it as a base model for pairwise comparisons. For an extensive survey of other connections and motivations, we recommend \cite{hamilton2025many}.

\subsection{General pairwise comparisons}\label{sec:gpc}
Interpreting the BT model as a generalized linear model allows for a unification of various pairwise comparison models in the literature  \citep{schauberger2019btllasso}. This idea was formulated by \cite{han2023general}. In the unified framework in \cite{han2023general}, the comparison outcome $Y_{ij}$ takes values in a symmetric set $\A \subseteq \mathbb{R}$. The symmetry of $\A$ reflects the reciprocity of comparison outcomes: the event that $i$ defeats $j$ is equivalent to $j$ losing to $i$. Common choices of $\A$ include $\A = \{-1, 1\}$ for binary outcomes, $\A = \{-1, 0, 1\}$ to include ties, and $\A =\R$ for cardinal outcomes. 

The probability mass function of $Y_{ij}$ (or the density if $Y_{ij}$ is continuous) is parameterized as $f(Y_{ij}; u_i - u_j)$ for some \emph{valid} function $f \colon \A \times \mathbb{R} \to \mathbb{R}_+$. By definition, $f$ is valid if it satisfies the symmetry condition $f(y; u) = f(-y; -u)$, is decreasing in $u$ for fixed $y < 0$, and is uniformly bounded. Important examples of valid functions include those that are strictly log-concave in $u$ for every $y \in \A$. Many popular pairwise comparison models can be placed within this framework by choosing an appropriate $f$, including the BT model, the Thurstone--Mosteller model \citep{thurstone1927law, mosteller1951remarks}, the Rao--Kupper \citep{rao1967ties} and Davidson \citep{davidson1970extending} models for ties, the more general ordinal models \citep{tutz1986bradley,agresti1992analysis} and the paired cardinal model \citep{shah2016estimation}. More recently, \citet{song2026semiparametric} studied a semiparametric pairwise comparison model in which the link function is left unspecified.

A crucial property of models defined by such valid $f$'s is the \emph{strong stochastic transitivity} (SST) \citep{joe1988majorization}. When $\A$ is discrete, this property reads
\begin{align*}
\min\{p_{ij}, p_{jk}\} \ge \frac{1}{2}\quad \implies\quad p_{ik} \ge \max\{p_{ij}, p_{jk}\},
\end{align*}
where $p_{ij}
\coloneqq \sum_{y > 0} f(y; u_i - u_j)+  \{f(0; u_i - u_j)/2\}$
is the probability that object $i$ defeats object $j$ with ties split equally. SST is a parsimonious and interpretable structural assumption and has also been adopted in various nonparametric extensions \citep{chatterjee2015matrix,chatterjee2019estimation}. In practice, the validity of the SST assumption can be assessed via hypothesis testing \citep{makur2025hypothesis,makur2025minimax}. However, the universal applicability of the SST assumption has been questioned. We discuss potential model misspecification under the SST assumption and related remedies in Section~\ref{sec:nonp}.

General pairwise comparison models are also closely related to synchronization problems, which seek to recover latent group elements from noisy pairwise relative transformations. The difference-based parametrization of these models can be viewed as synchronization over the additive group $\R$, where $u_i-u_j$ represents the relative displacement between objects $i$ and $j$ \citep{huang2017translation,araya2023dynamic}. The choice of $f$ then specifies the noise model for the corresponding synchronization problem. This viewpoint connects the general pairwise comparison models to synchronization over groups, whose other instances include phase synchronization \citep{singer2011angular,gao2021exact}, rotation synchronization \citep{boumal2013robust,gao2023optimal} and $\mathbb{Z}_2$ synchronization \citep{cucuringu2015synchronization}.

\subsection{Multiple comparisons}\label{sec:pl}
Unlike general pairwise comparison models, the PL model extends the BT model to multiple comparisons. For $2 \le k \le n$ and a subset of objects $e = \{i_1, \ldots, i_k\} \subseteq [n]$, the comparison outcome $Y_e\in \{i_{\pi(1)} \succ \cdots \succ i_{\pi(k)} : \pi \in S_k\}$ is a total order on $e$, where $S_k$ is the symmetric group on $[k]$, and $i_{\pi(j)}$ denotes the object in $e$ ranked in position $j$. In the PL model, $Y_e$ is modeled as a random variable with distribution
\begin{align}
\P(Y_e = i_{\pi(1)} \succ \cdots \succ i_{\pi(k)})
= \prod_{j = 1}^{k-1} \frac{\exp(u_{i_{\pi(j)}})}{\sum_{\ell = j}^{k} \exp(u_{i_{\pi(\ell)}})}.
\tag{PL}\label{eq:pl}
\end{align}
This probability is obtained by sequentially selecting objects from top to bottom; at each stage, an object is chosen from the remaining set with probability proportional to its strength (also called the multinomial logit model or Luce's choice model).

The defining property of the PL model is \emph{Luce's choice axiom} \citep{luce1959individual}, which states that the relative likelihood of selecting one object over another is unaffected by the presence or absence of additional objects. This axiom implies the \emph{internal consistency} of the model, a generalization of the axiom to multiple objects \citep{hunter2004mm}. Luce's choice axiom is a probabilistic formulation of the \emph{independence of irrelevant alternatives} axiom. While the latter is often criticized for being overly restrictive and leading to paradoxes such as the red bus--blue bus problem, it is frequently used due to mathematical tractability.

The PL model admits a latent-variable interpretation that is particularly convenient for Bayesian inference \citep{guiver2009bayesian, caron2012efficient}. If each object $i$ is associated with a latent variable $X_i = u_i + \varepsilon_i$, where $\varepsilon_i$ are i.i.d. $\mathrm{Gumbel}(0, 1)$, then the order statistics of these variables follow the PL distribution \citep{yellott1977relationship}. A similar interpretation based on exponential distributions for strengths can be found in \cite{diaconis1988group}. These interpretations put the PL model within a specific instance of random utility models \citep{mcfadden1972conditional}.

\subsection{Generalizations}\label{sec:gbt}

The utility vector of BT/PL models is static. In real-world applications, comparison data often exhibit heterogeneity across different contexts, populations, or time periods. This has motivated a family of generalizations that retain the utility-difference structure while expanding the modeling scope. Below, we mention two extensions along this direction: \textit{mixture models} and \textit{covariate-assisted models}.

\paragraph*{BT mixture models} A BT mixture model extends the BT model by assuming that the comparison probability between $i$ and $j$ results from a convex combination of $H$ latent BT models with utility vectors $\{\bm u^{(h)}\}_{h=1}^H$:
\begin{equation}
\P(Y_{ij} = i \succ j) = \sum_{h\in [H]} \omega_h \sigma(u_i^{(h)} - u_j^{(h)}),\tag{BTMixture}
\end{equation}
where $\omega_h > 0$ and $\sum_{h\in [H]} \omega_h = 1$. Similar formulations can be defined for the PL model and random utility models \citep{zhao2016learning}. While these models are practically appealing, they are notoriously difficult to analyze. Recent research has focused on the identifiability and efficient learning of these mixtures \citep{tkachenko2016plackett, liu2019learning, zhao2019learning, zhao2016learning, nguyen2023efficient, zhang2022identifiability}, as well as Bayesian extensions for more complex ranking structures \citep{mollica2017bayesian}.

\paragraph*{Covariate-assisted BT models} Covariate-assisted BT models integrate auxiliary feature information into utility modeling \citep{schauberger2019btllasso, yu2019analysis, bockenholt2006thurstonian}. One common approach utilizes linear regression, parameterizing utilities as a linear function of covariates in addition to intrinsic merits \citep{cattelan2012models, li2022bayesian, fan2024covariate, fan2024uncertainty, dong2025statistical}. A comprehensive modeling framework in this vein was introduced by \cite{schauberger2019btllasso} and studied in the PL setting under the name of the PL model with dynamic covariates (PlusDC) \citep{dong2025statistical}. In the pairwise setting, the PlusDC model can be stated as follows:  
\begin{align*}
\P( Y_{ij} = i \succ j )
&= \frac{\exp\lp u_i + \bm x_{\{i, j\}, i}^\top\bm v\rp}{\exp\lp u_i+\bm x_{\{i, j\}, i}^\top\bm v\rp + \exp\lp u_j+\bm x_{\{i, j\}, j}^\top\bm v\rp}\\
&= \sigma\lp u_i - u_j + (\bm x_{\{i, j\},i}-\bm x_{\{i, j\}, j})^\top\bm v\rp,\tag{PlusDC}\label{plusdc}
\end{align*}
where $\bm x_{\{i, j\}, i}\in \mathbb{R}^d$ is the covariate vector for object $i$ on edge $e=\{i, j\}$, and $\bm{v} \in \mathbb{R}^d$ is the global linear effect vector. This formulation strictly generalizes the BT model and encompasses several of its extensions as special cases, including the BT model with home-field advantage in sports analytics \citep{agresti1990categorical}. When the covariate is time, one may instead model the utility of object $i$ as a time-varying function $u_i(t)$ to capture temporal dynamics more flexibly \citep{fahrmeir1994dynamic,cattelan2013dynamic,bong2020nonparametric}.

While \eqref{plusdc} focuses on object and context features, it does not account for heterogeneous subject effects \citep{schauberger2019btllasso}. Such refinements can be valuable for modeling involving subject information. However, we omit this additional complexity in subsequent discussion, as many of the corresponding statistical foundations remain largely open.

Another important extension is to replace the linear covariate effect
$\bm x_{e,i}^{\top}\bm v$ with a shared nonlinear function of the
covariates, which may be parameterized by a neural network
\citep{luo2026deep}. This nonlinear formulation provides a natural
connection to preference learning for large language models (LLMs). In this setting, human feedback often consists of pairwise judgments of LLM-generated responses to the same prompt. A prominent example is reward modeling in reinforcement learning from human feedback (RLHF) \citep{christiano2017deep,ouyang2022training,bai2022training}. For each comparison $e$, two candidate responses $i$ and $j$ are generated for the same prompt. Here, the utility is modeled as a function of the prompt--response pair:
\begin{align}
\P(Y_{ij}=i\succ j)
=
\sigma\left(
r(\bm x_{e,i})-r(\bm x_{e,j})
\right),
\tag{BT-RM}\label{eq:bt_rm}
\end{align}
where $\bm x_{e,i}$ denotes the prompt--response pair associated with response $i$, and $r$ is the reward function to be learned. This agrees with the setup of \citet{luo2026deep} when the object-specific utility vector is set to zero. Learning the shared reward function enables preference predictions for previously unseen prompt--response pairs. Related approaches to LLM alignment include alternative link functions, classification-based objectives, and direct preference optimization methods \citep{siththaranjan2024distributional,sun2025rethinking,rafailov2023direct,azar2024general,ethayarajh2024model,meng2024simpo}.

BT models also arise in LLM evaluation, but with a different choice of objects. Here, the objects are the LLMs themselves, and each comparison records a preference between responses generated by two models to the same prompt \citep{zheng2023judging,chiang2024chatbot}. The goal is to estimate model-specific utilities that summarize relative performance across prompts. The standard BT model therefore provides a baseline for ranking LLMs, while extensions can incorporate richer comparison outcomes and covariate information \citep{ameli2025a,frick2025prompttoleaderboard,xu2026judge}.

\subsection{Limitations}\label{sec:nonp}

\color{black}
The pairwise comparison probabilities induced by the BT model satisfy SST and therefore admit a globally consistent latent ordering of the objects. This assumption is restrictive and fails in settings such as Rock--Paper--Scissors, where cyclic preferences are intrinsic. In such cases, no latent score vector can reproduce the underlying pairwise preference probabilities, and the BT model is therefore misspecified. See also \citet{munos2024nash,chakraborty2024maxminrlhf,Liu2026impossibility,pukdee2026what} for recent discussions of analogous representational limitations of BT-based reward models for LLMs.

This limitation of the BT model can be understood through the lens of degrees of freedom. A pairwise comparison probability matrix on $n$ objects has $n(n-1)/2$ free parameters. The BT model (and the general pairwise comparison models in Section~\ref{sec:gpc}) parametrizes this matrix as a nonlinear transformation of a gradient flow, reducing the number of free parameters to $n-1$. This reduction yields a parsimonious model but comes at the cost of expressive power. To address this limitation, recent work relaxes the gradient-flow constraint and estimates the pairwise comparison probability matrix using techniques such as matrix completion under alternative structural assumptions \citep{massimino2013one,levy2021ranking,chen2024note,zhang2025rank,lee2025pairwise}.

Alternatively, one can infer a ranking directly from pairwise comparison data. Representative examples include counting algorithms such as the Borda count \citep{shah2018simple} and its weighted variants \citep{wauthier2013efficient}. These methods aggregate observed pairwise outcomes locally and do not exploit indirect comparisons along paths in the comparison graph. Nevertheless, they admit computationally efficient and robust implementations.

\color{black}

\section{Data structure}\label{sec:3}

\subsection{Comparison data}

For generality, we formalize the data structure for the PL model. Comparisons among $n$ objects are represented by a hypergraph $\mathcal H_n(E)=([n],E)$ and outcomes $\{Y_e\}_{e\in E}$. Each $e\in E$ specifies the objects compared on one occasion. In practice, the collection $E$ can be a multiset, allowing repeated comparisons on the same subset. Pairwise data are the special case in which $|e|=2$ for every $e\in E$.

By fixing an ordering of the elements in $e = \{i_1, \ldots, i_k\}$ (e.g., $i_j < i_\ell$ for $j < \ell$), we can identify a full ranking on $e$ with an element of the symmetric group $S_k$:
\begin{align*}
\pi \in S_{k} \Longleftrightarrow i_{\pi(1)} \succ i_{\pi(2)} \succ \cdots \succ i_{\pi(k)},
\end{align*}
where $\pi(j)$ denotes the order of the object in $e$ ranked in position $j$. In some applications, only a partial ranking is observed. A common scenario is the \textit{choice-$s$} ranking ($s < k$): $i_{\pi(1)} \succ \cdots \succ i_{\pi(s)}\succ\text{others}$, where only the first $s$ positions are revealed. In such cases, the observed outcome corresponds to a subset of $S_k$, and these subsets form a partition of $S_k$.

In statistical analysis, we often assume that the outcomes $\{Y_e\}_{e \in E}$ are \textit{independent} conditional on $\mathcal{H}_n(E)$. This assumption implicitly assumes that the choice of which objects to compare is independent of their utilities/strengths. In real-world networks, however, the comparison structure and observed outcomes may be coupled. Capturing such dependencies requires the joint modeling of network structure and choice behavior \citep{de2018physical, kawakatsu2021emergence}.

\subsection{Comparison graphs}\label{subsec:comparison_graphs}

The structure of comparison graphs is fundamental to statistical analysis in BT/PL models. If $E$ contains all possible pairwise edges, $\HH_n(E)$ reduces to a complete graph on $n$ vertices. This represents the classic ``round-robin'' tournament setting \citep{zermelo1929berechnung} commonly found in sports analytics within a single season or league. However, when $n$ is large, obtaining comparisons for every pair becomes logistically expensive. Under such circumstances, it is typically assumed that $\mathcal{H}_n(E)$ is sparse but connected; the latter being a necessary condition to obtain a global ranking. This large-$n$ regime has become a focal point of recent research.

A popular approach to modeling graph sparsity is to assume that $\HH_n(E)$ is sampled from some random graph model, most notably the $\ER$ (ER) model \citep{MR0125031} or its hypergraph counterparts \citep{fan2025ranking}. Much of the modern asymptotic theory for BT/PL models relies on these sampling assumptions \citep{Negahban2012RankCR, chen2019spectral, han2020asymptotic, chen2022partial, gao2023uncertainty, fan2025ranking}. Nevertheless, the ER assumption can be too restrictive to capture certain characteristics of real-life networks, such as clustering and imbalanced degree distributions \citep{watts1998collective, barabasi1999emergence}. There is growing interest in alternative sampling frameworks, such as the (hypergraph) stochastic block model (SBM) or general topologies \citep{shah2016estimation, hendrickx2020minimax, li2022ell, bong2022generalized, han2023general, han2024statistical, yang2024top, han2025unified, fan2025spectral}.

\begin{figure}[ht!]
    \centering\vspace{-0.5cm}
    \includegraphics[width=1\linewidth]{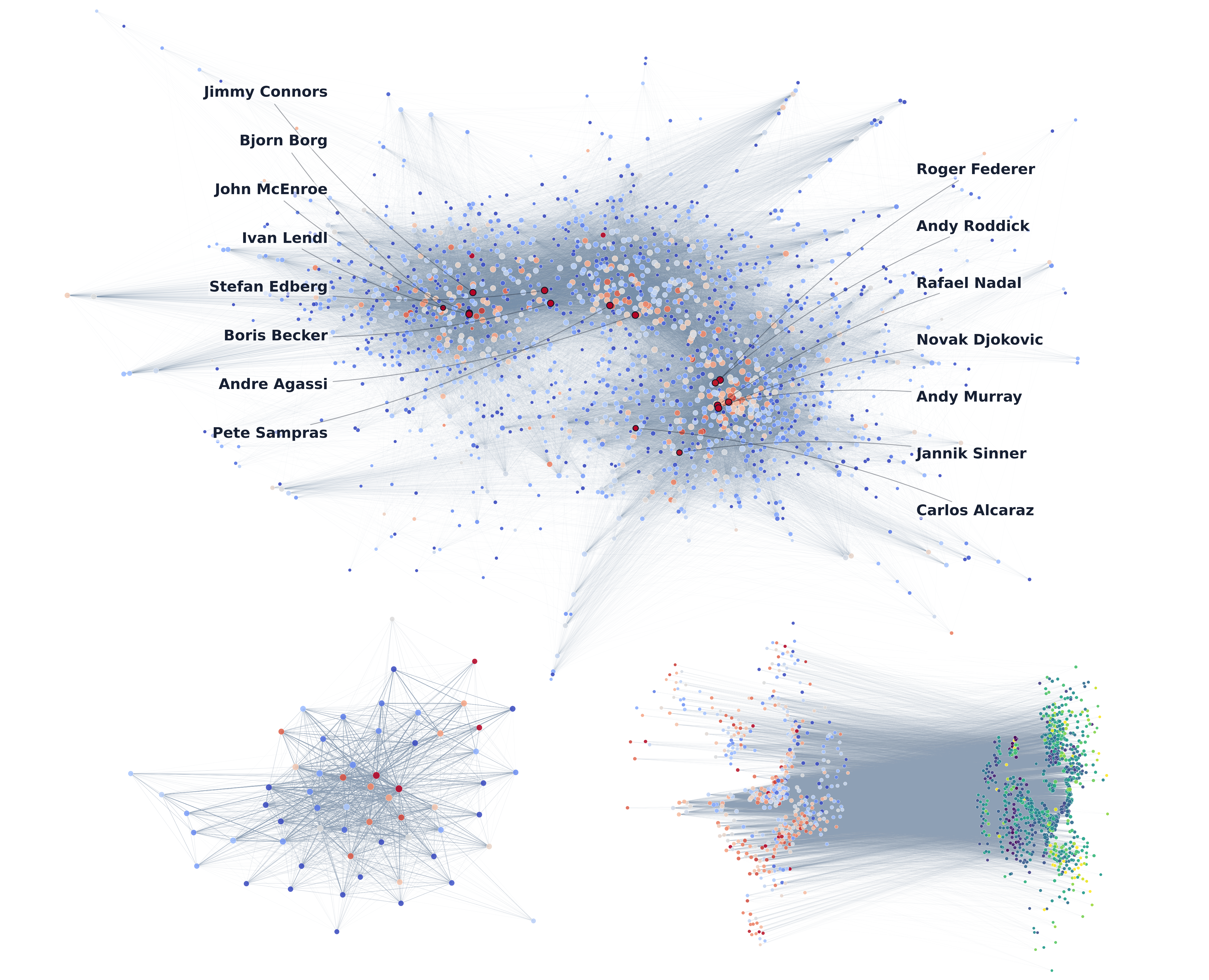}\vspace{-0.5cm}
    \caption{Comparison graph topologies from an extended ATP tennis dataset \citep{datasetATP} (top), an animal social network dataset \citep{datasetVervetMonkeys} (bottom left), and an educational evaluation dataset \citep{dataset_irt_kdd_algebra1} (bottom right). They exhibit temporally clustered, well-connected homogeneous, and bipartite structures, respectively.}
    \label{fig:graph_vis}
\end{figure}

For the PL model, another extreme setting is where each edge in $E$ contains all objects simultaneously. This scenario is of significant interest in social-choice applications like voting. In the large-$n$ regime, such comprehensive comparisons become less practical from a data-collection perspective. We recommend the monograph by \citet{xia2019learning} for recent advances along this thread.

\subsection{Covariates}\label{subsec:cov}
As discussed in Section~\ref{sec:gbt}, modern applications of BT/PL models often incorporate auxiliary information to enhance predictive performance and interpretability. These covariates can be categorized based on whether they characterize the objects themselves or the specific context of the comparison. To formalize this, let $\bm x_{e, i} \in \mathbb{R}^d$ denote the covariate vector associated with object $i$ within a specific edge $e\in E$ (assuming $i\in e$). We distinguish between two types of covariates:
\begin{itemize}
\item (Static covariates) $\bm x_{e, i} = \bm x_i$ for all $e \in E$. These are inherent properties of an object that remain constant across all comparisons, such as a player's physical attributes that do not change over time. 
\item (Dynamic covariates) $\bm x_{e, i}$ varies across different edges $e$. These are comparison-specific factors, such as the ``home-field advantage'' in sports or ``positional bias'' in a list of search results.  
\end{itemize}
A complete classification of covariates accounting for subject effects can be found in \cite{schauberger2019btllasso}. 
As we will see in Section~\ref{sec:ident}, dynamic covariates play a vital role in model identifiability and data fitting, allowing the model to distinguish between an object's intrinsic merit and the population effects.

\section{Statistical theory}\label{sec:st}

\subsection{Model identifiability}\label{sec:ident}

Since the winning probabilities in the BT model \eqref{eq:BT_prob} and PL model \eqref{eq:pl} depend only on utility differences, the models are invariant to a common shift, i.e., $\bm{u}$ and $\bm{u} + c\mathbf{1}$ (for any $c \in \mathbb{R}$) induce the same probability distribution on all edges. Consequently, the utility vector is not identifiable without additional constraints. While much of the literature assumes identifiability once a suitable linear constraint is imposed, the notion of identifiability itself is often not formally defined. 

Here, we define the identifiability of a BT/PL model as a property conditional on the underlying comparison graph. Specifically, we call a model identifiable if any two utility vectors $\bm{u}, \bm{u}'$ that induce the same probability distribution on $\mathcal{H}_n(E)$ satisfy $\bm{u} = \bm{u}'$. Under this definition, a linear constraint of type $\bm b^\top\bm u = 0$ with $\bm b^\top \bm 1\neq 0$ ensures the identifiability of the model provided that $\HH_n(E)$ is connected. This result also extends to the general pairwise comparison models \citep{han2023general}. Among such constraints, \citet{wu2022asymptotic} recommends using $\bm 1^\top \bm u = 0$ due to its variance-minimization property in the BT setting.

For covariate-assisted BT/PL models, the notion of identifiability can be modified to be conditional on both the comparison graph and covariates. Under this definition, the PlusDC model in \eqref{plusdc} is identifiable under a minimal constraint such as $\bm 1^\top \bm u = 0$, assuming a maximum-rank condition on an augmented matrix involving the incidence structure of the comparison graph and covariate differences \citep{dong2025statistical}. This condition effectively requires all covariates to be in generic positions or sufficiently random; all static covariates can be absorbed into the object utilities and additional constraints are needed to ensure identifiability \citep{fan2024uncertainty}. This observation underscores the importance of using dynamic covariates to improve model fit.

Identifiability becomes more subtle for BT/PL mixture models. Following the standard definition for mixture models (e.g., \citet{allman2009identifiability}), strict identifiability requires uniqueness at every admissible parameter value up to label switching, whereas generic identifiability allows an exceptional set of Lebesgue measure zero. Thus, a model may be generically identifiable but not strictly identifiable.
For BT/PL mixture models, aside from the inherent label-switching ambiguity, a more fundamental identifiability issue arises due to the dependence structure of the mixture components. The latter issue for the PL mixture model was first addressed by \citet{zhao2016learning}. When each edge contains all $n$ objects, \cite{zhao2016learning} showed that the PL mixture model is non-identifiable if the number of components satisfies $H \geq (n+1)/2$, while it is generically identifiable if $1 \leq H \leq \lfloor (n-2)/2 \rfloor!$ and $n \geq 6$. These results were later generalized to partial observation settings and structured comparison graphs \citep{zhao2019learning}. In particular, \cite{zhao2019learning} proved that the BT mixture model is \textit{not} identifiable via an explicit construction. Despite this negative result, \citet{zhang2022identifiability} showed that a BT mixture model with $H=2$ components is generically identifiable on a complete graph $\mathcal{H}_n$ if $n \geq 5$ using algebraic geometry techniques. This result addresses a long-standing gap in the field, but many more remain open. See \cite{zhang2022identifiability} for a comprehensive review of related work.


\subsection{Estimation}\label{sec:est}

Parameter estimation is a crucial step when applying the BT model to fit data. This section collects several commonly used approaches for parameter estimation in the BT/PL models and their variants. We do not cover mixture models, as their methods are mostly empirical due to the nonconcave objective; an algorithmic discussion on them is deferred to Section~\ref{sec:alg}.

\paragraph*{Likelihood-based estimators.} The standard parametric approach is \textit{maximum likelihood estimation} (MLE). Assuming the comparison graph $\HH_n$ is simple and the comparison outcomes $\{Y_e\}_{e\in E}$ are independent, the MLE for the true utility vector $\bm u^*$ with $\bm 1^\top \bm u^* = 0$ (conditioned on $\mathcal{H}_n$) is given by
\begin{align}
\widehat{\bm u} = \argmax_{\bm u\in\mathbb{R}^n, \bm 1^\top\bm u = 0}\sum_{\{i, j\}\in E}\log \sigma(Y_{ij} (u_i - u_j)). \tag{BT-MLE}\label{btmle}
\end{align}
Here, we assume $Y_{ij}\in\{-1, 1\}$ following the general pairwise comparison notation in Section~\ref{sec:gpc} (with the corresponding $f(y;u)=\sigma(yu)$); other works often use $Y_{ij}\in\{0, 1\}$ according to the convention of logistic regression.

The constraint $\bm 1^\top\bm u = 0$ in \eqref{btmle} ensures the model is identifiable provided that $\mathcal{H}_n$ is connected. However, this does not guarantee the existence of a finite solution. The existence of the MLE is a probabilistic event that depends on the comparison outcomes $\{Y_e\}_{e\in E}$. For example, if object $i$ defeats every opponent it faces, its estimated utility $\widehat{u}_i$ will diverge to infinity because the likelihood is monotonically increasing in $u_i$. 

A classical result by \citet{MR0097876} says that the MLE uniquely exists and is finite if and only if the comparison graph, together with comparison outcomes, is \textit{strongly connected}: for every partition of the objects $[n] = I \cup I^\complement$, there exists an object in $I$ that defeats some object in $I^\complement$. This condition is equivalent to irreducibility when wins and losses are interpreted as state transitions. It implies that the negative objective in \eqref{btmle} is strictly convex and coercive over the feasible set. In the asymptotic regimes ($n \to \infty$), this condition holds with high probability provided the comparison graph is well-connected \citep{han2023general}. Similar existence results have been established for the PL model \citep{hunter2004mm}, general pairwise comparison models \citep{han2023general}, and the PlusDC model \citep{dong2025statistical}. Alternatively, one may add regularization (e.g., ridge) to ensure the unique existence of a solution regardless of the comparison outcomes \citep{chen2019spectral, glickman2026regularization}.

For PL models involving partial rankings, the MLE in \eqref{btmle} can be adapted into a marginal-MLE by considering the marginal likelihood of the observed partial orders. Aside from MLE/marginal-MLE, one can also use rank-breaking techniques to decompose multiple comparisons into pairwise comparisons \citep{azari2013generalized}. For instance, a full breaking of $\{2\succ 3\succ\{1, 4\}\}$ yields $\{2\succ 3\}, \{2\succ 1\}, \{2\succ 4\}, \{3\succ 1\}, \{3\succ 4\}$. While the pairwise comparisons obtained in this way may be dependent, one can treat them as independent observations arising from a (misspecified) BT model. This approach yields an estimator that can be interpreted as a generalized moment estimator \citep{azari2013generalized} or quasi-MLE \citep{han2025unified}. We refer the reader to \cite{han2025unified} for a unified perspective on the likelihood-based estimators in the PL model. 

\paragraph*{Spectral methods.} Spectral methods have a long history in rank aggregation \citep{kendall1955further, keener1993perron, dwork2001rank}, with the PageRank algorithm being a cornerstone example \citep{brin1998anatomy}. The core idea is to construct an ergodic Markov chain on the set of objects such that its stationary distribution reveals the relative rank information. To connect spectral ranking with the BT model, \citet{Negahban2012RankCR} defined an ergodic Markov chain on $[n]$, with transition probabilities from $i$ to $j$ proportional to the probability that $j$ defeats $i$ for $\{i, j\}\in E$. This Markov chain satisfies the detailed balance condition, and its stationary distribution is proportional to the strength vector $\bm\gamma$. In practice, the transition matrix can be estimated empirically and used to compute the stationary distribution, yielding the spectral algorithm known as \textit{RankCentrality} (RC).


Similar ideas were extended to the PL model with choice-one observations by interpreting the estimating equations of the MLE as the global balance equations of a continuous-time Markov chain \citep{maystre2015fast}. Specifically, they propose \textit{Luce Spectral Ranking} (LSR), a single-step spectral estimator obtained from the stationary distribution of a Markov chain, and its iterative variant, Iterative LSR (I-LSR), which repeatedly updates the transition rates and converges to the MLE. Within this framework, RC can be viewed as a special case of LSR in the pairwise setting. A different spectral method, called Accelerated Spectral Ranking (ASR), was proposed by \cite{agarwal2018accelerated}, motivated by the observation that the discrete-time chain associated with LSR can have large self-loop probabilities and hence mix slowly. ASR constructs a faster-mixing chain while retaining the ability to recover the same parameter estimate through a linear transformation of its stationary distribution. PL models with general comparisons can be reduced to the choice-one setting via rank-breaking techniques \citep{azari2013generalized}. An in-depth analysis of the estimation efficiency of spectral methods relative to the MLE is provided in \cite{chen2022optimal, fan2025spectral}.

\paragraph*{Bayesian approach.} One drawback of the MLE is its potential non-existence when the comparison graph does not satisfy the strong connectivity condition. Bayesian methods address this by incorporating prior information in the model. By Bayes' theorem, the posterior distribution of the utility vector $\bm u$ given the comparison data $ \{Y_e\}_{e\in E}$ satisfies 
\begin{align}
p(\bm u \mid \{Y_e\}_{e\in E})\propto g(\bm u)\times\prod_{\{i, j\}\in E}\sigma(Y_{ij} (u_i - u_j)),\tag{BT-Bayes}\label{BT-Bayes}
\end{align}
where $g(\cdot)$ is the prior density on $\bm u$. A point estimate for $\bm u$ can be obtained by computing the posterior mean of $p(\bm u \mid \{Y_e\}_{e\in E})$ or the maximum a posteriori (MAP) estimate. 

Different priors have been proposed for the BT model, including conjugate-family priors on $\bm\gamma$ \citep{davidson1973bayesian}, Gaussian distributions on $\bm u$ \citep{leonard1977alternative}, and Dirichlet distributions on $\bm\gamma$ \citep{chen1984bayes}. There is a one-to-one correspondence between priors on $\bm u$ and $\bm\gamma$, and different priors may induce equivalent model distributions. These issues were systematically studied by \cite{whelan2017prior}, who also compared various priors based on certain desiderata. 

For the PL model, the latent-variable interpretations discussed in Section~\ref{sec:pl} suggest the use of independent gamma priors on the score vector $\bm\gamma$. \citet{guiver2009bayesian} addressed estimation and inference in this setting using Expectation Propagation techniques \citep{minka2004power} to approximate the posterior; \citet{caron2012efficient} introduced a data augmentation algorithm that enables exact posterior sampling via Gibbs samplers. The Bayesian approach was later extended to the infinite-object setting \citep{caron2014bayesian} and models with auxiliary covariate information \citep{li2022bayesian}. 

From the modeling perspective, the Bayesian approach is similar to regularization (i.e., the MAP estimate is a regularized MLE per se). We will discuss this in further detail when reviewing the algorithms for BT/PL models in Section~\ref{sec:alg}.

\subsection{Consistency}

In this section, we review consistency results for the MLE in BT/PL models. Since many spectral methods can be viewed as proxies for the MLE, their consistency results often follow analogously. We are interested in the regime $n \to \infty$. In this setting, the number of parameters grows along with the number of observations, so classical large-sample theory for the MLE does not directly apply \citep{neyman1948consistent}. On the other hand, the norm used to measure the estimation error matters, as different norms may not be asymptotically equivalent. Some common choices include:
\begin{itemize}
\item (Normalized $\ell_2$ norm) $\|\widehat{\bm u} - \bm u^*\|_2/\sqrt{n}$;
\item (Laplacian semi-norm) $\|\widehat{\bm u} - \bm u^*\|_\mathcal{L} \coloneqq \sqrt{\langle\widehat{\bm u} - \bm u^*, \mathcal{L}(\widehat{\bm u} - \bm u^*)\rangle}$, where $\mathcal{L}$ is the graph Laplacian associated with $\mathcal{H}_n$ normalized by $n$ (for hypergraphs, see \cite[Eq. (10)]{shah2016estimation});
\item ($\ell_\infty$ norm) $\|\widehat{\bm u} - \bm u^*\|_\infty$.
\end{itemize}
Since the $\ell_\infty$ norm is not normalized by $n$, convergence with respect to it implies \textit{uniform consistency}. This property allows the estimated utilities to be transferred to object rankings under suitable separation conditions. Uniform consistency is stronger than normalized $\ell_2$ convergence and is more difficult to prove. The Laplacian semi-norm is often used to obtain cleaner results than the normalized $\ell_2$ norm by internalizing the structure of the comparison graph. In the rest of this section, we focus on recent advances in the uniform consistency of the MLE and briefly mention other relevant results when appropriate.

Uniform consistency depends critically on the topology of the comparison graph. Much of the theoretical work focuses on the homogeneous setting, where $\HH_n$ is sampled from an ER model $G(n, p_n)$. \cite{MR1724040} first established the uniform consistency of the MLE in the BT model for complete graphs ($p_n = 1$). This was later extended to the sparse regime $p_n \gtrsim (\log n)^3/n$ by \cite{han2020asymptotic}. This result is optimal up to polylogarithmic factors relative to the ER connectivity threshold $(\log n)/n$.

By a leave-one-out perturbation argument, also known as the cavity method in statistical mechanics, which constructs a sequence of auxiliary models by removing one object together with its associated edges, the requirement on $p_n$ can be sharpened to the optimal rate $p_n \gtrsim (\log n)/n$ for both the MLE \citep{chen2022partial} and the regularized MLE \citep{chen2019spectral}. Moreover, \cite{chen2019spectral} established uniform consistency for the RC algorithm. The leave-one-out method is particularly effective in homogeneous models and yields sharp results for estimation and ranking efficiency \citep{chen2022partial,chen2022optimal}. More broadly, it is an important tool in the analysis of synchronization problems, particularly for obtaining sharp coordinate-wise guarantees for spectral and iterative estimators \citep{ling2022improved,ling2022near}. These results for the BT model have also been extended to multiple comparisons on homogeneous hypergraphs.

Although homogeneous graph assumptions appear in many existing results, they are often inappropriate for modeling real-world networks. Generalizing results from homogeneous to heterogeneous settings has attracted growing interest. Earlier works by \citet{yan2012sparse} and \citet{li2022ell} revealed connections between uniform consistency and comparison graph topology, though their bounds were sub-optimal when specialized to the homogeneous case. Subsequent research has developed new machinery to improve these results, including graph-based chaining \citep{han2023general}, preconditioned gradient descent (PGD) analysis \citep{chen2023ranking, yang2024top}, and leave-one-out analysis combined with reweighting \citep{fan2025spectral}. We briefly review the ideas behind these recent developments. 

The graph-based chaining technique is a general approach for relating estimation error to graph topology. Roughly speaking, it chains objects with extreme estimation errors through a nested sequence of vertices along which the errors vary smoothly. The degree of smoothness depends on the unnormalized Cheeger constant \citep{MR1421568}, and the sequence length is uniformly bounded by the diameter of the admissible subsequences, a generalized notion of graph diameter. This allows $\|\widehat{\bm u} - \bm u^*\|_\infty$ to be controlled by the connectivity of $\HH_n$. Such a result enables sharp analysis of clustered random graph models, such as SBMs with near-minimal connectivity. Moreover, it can be extended to other frameworks, including the general pairwise comparison models \citep{han2023general}, the PL model \citep{han2025unified}, and the PlusDC model \citep{dong2025statistical}.

In contrast, \citet{chen2023ranking} analyzed entrywise error using a PGD algorithm initialized at the ground truth. The results involve some complicated terms, which were simplified by \citet{yang2024top} using tools from spectral graph theory. Specifically, \citet{yang2024top} showed that the entrywise errors for the MLE can be uniformly controlled by the maximum degree and spectral gap, or the Cheeger constant, via Cheeger's inequality \citep{MR1421568}. Although the Cheeger constant is better suited than its unnormalized counterpart for characterizing graph topology, it loses certain properties that may be crucial when dealing with heterogeneous structures. For instance, it is not monotone as edges are added to the graph. \citet{fan2025spectral} studied the spectral method for general multiple comparisons on fixed comparison graphs. Under the BTL or PL model and suitable regularity conditions, they showed that a two-step spectral estimator with appropriately estimated weights can achieve the same asymptotic efficiency as the MLE. This result bridges the gap between computationally efficient spectral methods and statistically optimal likelihood-based estimation.

We briefly mention some consistency results in other norms. \citet{Negahban2012RankCR} bounded the normalized $\ell_2$ error for RC via the ratio of the maximum and minimum degrees, which is tight in the ER setting under the optimal sparsity condition $p_n \gtrsim (\log n)/n$. \citet{shah2016estimation} generalized this result to general graph topologies and established minimax bounds for both the normalized $\ell_2$ and Laplacian norms; see also \citet{hajek2014minimax}. Other works on the uniform consistency of the MLE also contain results for the normalized $\ell_2$ norm \citep{chen2019spectral, chen2022partial, fan2024uncertainty}, which are not surveyed here. More recently, consistency guarantees have also been studied for covariate-assisted BT models \citep{fan2024uncertainty, dong2025statistical} and BT-type reward models in RLHF \citep{zhu2023principled,zhan2024provable,liu2025uncertainty,luo2026learning}.

\subsection{Asymptotic normality}

Asymptotic normality is the next step once uniform consistency is established. This section surveys the asymptotic normality results of the MLE in BT/PL models and their applications in statistical inference, such as constructing confidence intervals and hypothesis tests.

Similar to consistency results, early work on asymptotic normality focused on the homogeneous setting. \cite{MR1724040} obtained asymptotic normality for the MLE in the BT model with a complete comparison graph by leveraging an explicit matrix formula to approximate the inverse Hessian. This approach was later adapted for sparser graphs \citep{yan2012sparse, han2020asymptotic}. \cite{liu2023lagrangian} proposed a different debiasing technique to prove asymptotic normality in the ER model with optimal sparsity $p_n \gtrsim {(\log n)}/{n}$, though their result requires each edge to contain at least $(n \log n)^2$ repeated comparisons.

A breakthrough came from \cite{gao2023uncertainty}, which established the first optimal asymptotic normality result for both the MLE and the spectral estimator associated with the RC algorithm. In the ER setting, this was achieved under the sparsity assumption $p_n \gtrsim {\text{poly}(\log n)}/{n}$ via a leave-two-out method. This technique generalizes the leave-one-out method and is effective for disentangling dependencies among random variables in the analysis. It was subsequently applied to establish the asymptotic normality of the MLE for multiple comparisons \citep{fan2025ranking}. 
Establishing asymptotic normality under general graph topologies is significantly more challenging. For the PL model with deterministic comparison graphs, \cite{han2025unified} provided the first asymptotic results for a class of likelihood-based estimators in the PL model using a truncated error analysis. The key insight was to characterize the Fisher information matrix as a weighted graph Laplacian and analyze it using spectral graph techniques. Under the ER assumption, their results are comparable to those of \cite{gao2023uncertainty, fan2025ranking}. However, they further allow the degree distribution to be heterogeneous, provided it remains balanced (i.e., the ratio between the maximum and minimum degrees can diverge with $n$ in a controlled manner). Compared to the consistency results, such a balancing condition is typically necessary in the absence of debiasing, as the convergence rate of low-degree objects may decelerate the faster convergence of high-degree ones. This method was further generalized to general pairwise comparison models \citep{han2024statistical}. In a separate line of work, \cite{fan2025spectral} established asymptotic normality for a weighted spectral estimator, also under general comparison graphs. See Table~\ref{tab:combined} for a summary of both the consistency and asymptotic normality results. 

{\color{black} Asymptotic normality, together with consistent estimation of the asymptotic variance, provides a basis for constructing confidence intervals for low-dimensional utility parameters.
However, inferential questions extend beyond uncertainty quantification for individual utilities and do not necessarily
follow directly from asymptotic normality of their estimators. For example, hypothesis testing for high-dimensional utility parameters can be addressed through the likelihood-ratio framework of \citet{yan2025likelihood}.
}
A related but more challenging problem is to perform inference directly on ranking positions \citep{liu2023lagrangian, gao2023uncertainty}. For this task, the Gaussian multiplier bootstrap has emerged as an effective method for rank-based inference \citep{fan2024covariate, fan2024uncertainty, fan2025ranking, fan2025spectral}. Recently, the repro samples method \citep{chandra2025finite} has been proposed as an alternative for ranking inference without using asymptotic normality.

\begin{table}[htb!]
\centering
\caption{Summary of theoretical results.}
\label{tab:combined}
\resizebox{\textwidth}{!}{%
\begin{tabular}{@{}l l l l l@{}}
\toprule
\textbf{Work} & \textbf{Model} & \textbf{Estimation} & \textbf{Techniques} & \textbf{Graph} \\
\midrule
\multicolumn{5}{l}{\textit{Uniform Consistency}} \\
\cmidrule{1-5}
 \cite{chen2019spectral} & BT & RMLE/SE & leave-one-out & Homogeneous \\
 \cite{chen2022partial} & BT & MLE & leave-one-out & Homogeneous \\
\cite{han2023general} & General Pairwise & MLE & Graph chaining & Heterogeneous \\
\cite{yang2024top} & BT & MLE & PGD & Heterogeneous \\
\midrule
\multicolumn{5}{l}{\textit{Asymptotic Normality}} \\
\cmidrule{1-5}
\cite{gao2023uncertainty} & BT & MLE & leave-two-out & Homogeneous \\
\cite{fan2025ranking} & multiple & MLE & leave-two-out & Homogeneous \\
\cite{han2025unified} & PL & MLE/QMLE & Truncated error analysis & Heterogeneous \\
\cite{fan2025spectral} & PL & WSE & leave-one-out & Heterogeneous \\
\bottomrule
\end{tabular}
}

\smallskip
\footnotesize
 RMLE: Regularized MLE; SE: Spectral Estimator; QMLE: Quasi-MLE; WSE: Weighted SE.
\end{table}

\section{Algorithms}\label{sec:alg}

This section first surveys algorithms for MLE and spectral methods. Then we discuss recent algorithmic advances for the Bayesian approach and mixture models.

\subsection{Iterative algorithms for the MLE}\label{subsec:mle}

We focus on the BT model and point out extensions to more general settings as needed. For ease of interpretation, we switch to the strength parameters $\gamma_i = \exp(u_i)$ when convenient.

\paragraph*{Fixed-point iteration (FPI) for MLE.} For $i, j\in [n]$, let $w_{ij}$ denote the number of times $i$ defeats $j$, and $n_{ij} = w_{ij} + w_{ji}$. The log-likelihood function in \eqref{btmle} can be written as $l(\bm \gamma) = \sum_{i, j\in [n]}w_{ij}\log(\gamma_i/({\gamma_i + \gamma_j}))$. Under the strong connectivity assumption, $l(\bm\gamma)$ has a unique maximizer satisfying $\prod_{i\in [n]}\gamma_i = 1$.

By the first-order optimality condition of the MLE, $\widehat{\bm\gamma}\coloneqq\exp(\widehat{\bm u})$ satisfies, for all $i\in [n]$,  
\begin{align}
\partial_i l(\widehat{\bm \gamma}) = 0 &\iff \sum_{j\in [n]}w_{ij} - \sum_{j\in [n]}\frac{n_{ij}\widehat{\gamma}_i}{\widehat{\gamma}_i + \widehat{\gamma}_j} = 0\label{e1}\\
&\iff \sum_{j\in [n]}\frac{w_{ij}\widehat{\gamma}_j}{\widehat{\gamma}_i + \widehat{\gamma}_j} - \sum_{j\in [n]}\frac{w_{j i}\widehat{\gamma}_i}{\widehat{\gamma}_i + \widehat{\gamma}_j} = 0. \label{e2}
\end{align}
Isolating $\widehat{\gamma}_i$ in \eqref{e1} yields the following FPI for $\widehat{\bm \gamma}$: 
\begin{align}
\gamma_i = \frac{\sum_{j\in [n]}w_{ij}}{\sum_{j\in [n]}n_{ij}/(\gamma_i + \gamma_j)}.\tag{Zermelo}\label{e3}
\end{align}
Normalization is needed after each iteration to enforce $\prod_{i\in [n]}\gamma_i = 1$. This algorithm, known as Zermelo's algorithm, was used by \cite{zermelo1929berechnung} in his original paper. \citet{hunter2004mm} extended this to other general pairwise comparison models and the PL model and analyzed its convergence behavior using the theory of Minorization--Maximization (MM). \cite{qu2025sinkhorn} linked \eqref{e3} to the Sinkhorn algorithm \citep{sinkhorn1964relationship} by viewing \eqref{e1} as a matrix balancing problem. 

Although widely used, Zermelo's algorithm may converge slowly in certain circumstances. \citet{NewmanJMLR2023} proposed an alternative FPI based on a different algebraic rearrangement of \eqref{e2}:
\begin{align}
\gamma_i = \frac{\sum_{j\in [n]}w_{ij}\gamma_j/(\gamma_i + \gamma_j)}{\sum_{j\in [n]}w_{j i}/(\gamma_i + \gamma_j)}\tag{Newman},\label{e4}
\end{align}
which was observed to converge substantially faster than \eqref{e3} across a variety of settings. \citet{han2026newman} established a local convergence theory for \eqref{e4} and showed that its acceleration over \eqref{e3} critically depends on the use of asynchronous updates. In particular, while the asynchronous implementation is locally convergent, the synchronous implementation of \eqref{e4} may diverge when the comparison graph is near-bipartite. In contrast, the convergence and convergence rate of Zermelo's algorithm are less sensitive to the choice between synchronous and asynchronous updates. A subsequent extension of \eqref{e4} to the PL model was proposed by \citet{yeung2025efficient}.

\paragraph*{Gradient-based optimization.} 
Since the BT model is a special case of logistic regression subject to an identifiability constraint under the utility parametrization $\bm u$, a standard approach is to apply projected gradient descent to $-l(\bm u)$ as
\begin{align}
\bm u^{(t+1)}=\Pi_{\bm 1^\perp}\left\{\bm u^{(t)}+\eta_t\nabla l(\bm u^{(t)})\right\},\tag{GD}\label{gd}\end{align}
where $\Pi_{\bm 1^\perp}$ enforces the sum-to-zero constraint. Since $-l(\bm u)$ is convex and smooth in $\boldsymbol{u}$, gradient descent with a properly chosen step size converges to the MLE whenever the MLE exists, with a local linear rate governed by the conditioning of the Hessian. \citet{vojnovic2023accelerated} established tight convergence-rate bounds for gradient descent and Zermelo's MM algorithm under the utility parametrization and showed that their linear convergence rates agree up to constant factors. This theoretical equivalence does not imply identical empirical performance, which may vary across problem instances; see Section~\ref{app:bt-optimization} in the Supplementary Material for an empirical comparison of the MM and Newman iterations.
For large-scale applications, stochastic gradient descent methods are commonly used. For example, the classical Elo update can be interpreted as an online stochastic gradient step for maximizing the BT log-likelihood \citep{kiraly2017modelling, olesker2024analysis}.

\paragraph*{Spectral methods.} We introduce spectral algorithms following the framework in \cite{maystre2015fast}, which points out a natural connection to the MLE. Compared to \eqref{e1}, \eqref{e2} groups the ``win'' and ``loss'' terms of $i$ separately, and this allows for a natural Markov-chain interpretation. Denoting $\bm Q_{ij}(\bm\gamma) = w_{ji}/(\gamma_i + \gamma_j)$ for $i\neq j$ and $\bm Q_{ii}(\bm\gamma) = -\sum_{j\neq i}\bm Q_{ij}(\bm\gamma)$, $\widehat{\bm\gamma}$ is proportional to a stationary distribution of a continuous-time Markov chain with generator matrix $\bm Q(\widehat{\bm\gamma})$,
\begin{align}
\eqref{e2}\iff\widehat{\bm\gamma}^\top\bm Q(\widehat{\bm\gamma}) = 0. \label{e5}
\end{align} 
Equation \eqref{e5} can be solved via nonlinear power iteration; that is, one first fixes $\bm\gamma$ in $\bm Q(\bm\gamma)$, solves \eqref{e5} to update $\bm\gamma$ under the normalization constraint, and repeats until convergence. With uniform initialization and suitable comparison graph design, this full algorithm recovers the I-LSR algorithm \citep{maystre2015fast} in the pairwise setting. The first iteration of I-LSR, called LSR, coincides with the RC algorithm in \cite{Negahban2012RankCR}. \citet{agarwal2018accelerated} developed an accelerated spectral ranking algorithm by considering a rescaled Markov chain compared to LSR. 

\subsection{Bayesian approach}
This section reviews the algorithms for the Bayesian method in the BT model. We first discuss how priors can be transformed and compared across different parameterizations. We then review algorithms for MAP estimation and posterior mean estimation.

\paragraph*{Prior comparisons.} Aside from the strength and utility parameterizations, the BT model can also be expressed using pairwise winning probabilities $\theta_{ij} = \gamma_i/(\gamma_i + \gamma_j)$ or log-odds ratios $\lambda_{ij} = u_i - u_j$. Similar to the identifiability issues for the MLE, different parametrizations of $\bm u$ or $\bm \gamma$ may yield the same distributions of $\bm\theta$ or $\bm\lambda$. To address this, \citet{whelan2017prior} introduced a parametrization approach based on an identifiable basis $\bm\lambda \in\mathbb R^{n-1}$ of log-odds ratios, such as $\lambda_i \coloneqq \lambda_{in}=u_i-u_n$ for $i\in [n-1]$. \citet{whelan2017prior} also proposed four desiderata for evaluating priors and highlighted two that satisfy these criteria: the Gaussian prior, whose MAP corresponds to MLE with $\ell_2$ regularization; and the beta-separable prior, whose MAP is equivalent to MLE on a dataset augmented by fictitious data.
Other commonly used priors, such as Haldane-like, maximum-entropy, and Dirichlet, fail at least one desideratum. \citet{phelan2018hierarchical} noted that the beta-separable prior expressed in $\bm u$-parameterization visually resembles the Gaussian prior.

\paragraph*{Algorithms for MAP.} 
Under the $\bm\gamma$-parameterization, the MAP estimator $\widehat{\bm\gamma}_{\rm MAP}=\argmax_{\bm\gamma}\{ l(\bm\gamma)+\log g(\log(\bm\gamma))\}$ can be seen as a regularized MLE. 
Early approaches used iterative algorithms analogous to classical MLE solvers, such as fixed-point updates in \citet{davidson1973bayesian} and Newton–Raphson optimization in \citet{leonard1977alternative}. \citet{caron2012efficient} developed an efficient data augmentation scheme using independent gamma priors and obtained the closed-form update as
\begin{align}
    \gamma_i = \frac{a - 1 + \sum_{j \in [n]} w_{ij}}{b + \sum_{j\in [n]} \{n_{ij}/(\gamma_i + \gamma_j)\}}.\tag{EM-MAP}\label{EM-MAP}
\end{align}
This update rule in \eqref{EM-MAP} contains Zermelo's algorithm as a special instance (setting $a=1$ and $b=0$) and extends to other generalizations \citep{caron2012efficient}.

\paragraph*{Algorithms for the posterior mean.} 
 Posterior mean estimation typically relies on either deterministic approximation or Markov chain Monte Carlo (MCMC) sampling. For deterministic approximation, \citet{guiver2009bayesian} applied Power Expectation Propagation to approximate the true posterior as a fully factorized product of unnormalized gamma distributions: 
$p(\bm \gamma \mid \{Y_e\}_{e\in E}) \approx \prod_{i=1}^n q_i(\gamma_i; \alpha_i, \beta_i)$, where $q_i(\gamma_i; \alpha_i, \beta_i) = \gamma_i^{\alpha_i-1}e^{-\beta_i \gamma_i}$. 
The shape $\alpha_i$ and rate $\beta_i$ parameters are iteratively updated through the message-passing algorithm. Alternatively, MCMC methods avoid functional approximations by estimating them through sampling. Early implementations relied on Metropolis--Hastings proposals \citep{adams2005bayesian,gormley2009grade}, while more efficient MCMC algorithms have been developed, such as the Gibbs samplers derived by \citet{caron2012efficient} and the Hamiltonian Monte Carlo adopted by \citet{phelan2018hierarchical}.

\subsection{Mixture models}

Most algorithms for the BT mixture model are developed for the PL mixture model. In this case, we seek to estimate parameters $\boldsymbol{\Theta} = \{(\omega_h, \boldsymbol{\gamma}^{(h)})\}_{h=1}^H$, where $\omega_h$ are the mixing weights (with $\sum_{h\in [H]} \omega_h = 1$) and $\boldsymbol{\gamma}^{(h)}$ is the strength vector for the $h$-th component. The E-step is standard and computes the posterior responsibilities $\tau_{eh}$ belonging to each component. The M-step updates $\{\boldsymbol{\gamma}^{(h)}\}_{h=1}^{H}$ by maximizing the joint log-likelihood function
\begin{align}
    \{{\boldsymbol{\gamma}}^{(h)}\}_{h\in [H]} =  \operatorname*{argmax}_{\{\boldsymbol{\gamma}^{(h)}\}_{h=1}^{H}} \sum_{h\in [H]}\sum_{e \in E} \tau_{eh} \log \mathbb{P}(Y_e \mid \boldsymbol{\gamma}^{(h)}), \label{M_ll}
\end{align}
where $\mathbb{P}(Y_e \mid \boldsymbol{\gamma}^{(h)})$ is the probability of $Y_e$ following the PL model with strength vector $\boldsymbol{\gamma}^{(h)}$.
\citet{gormley2008exploring} applied the MM algorithm to the M-step, yielding the weighted FPI,
\begin{align*}
    \gamma_i^{(h)} \leftarrow \frac{\sum_{e\in E} \tau_{eh} \mathbb{I}(i \in Y_e)}{\sum_{e\in E} \tau_{eh} \sum_{r: i \in {Y_e}{(r)}} \left(\sum_{j \in {Y_e}{(r)}} \gamma_j^{(h)}\right)^{-1}}, 
\end{align*}
where ${Y_e}{(r)}$ denotes the set of objects ranked at or below position $r$ in $Y_e$. Recently, \citet{nguyen2023efficient} proposed an efficient EM-LSR framework that integrates posterior responsibilities $\tau_{eh}$ into the transition matrix, complemented by a two-stage spectral initialization involving clustering and least squares. Alternative M-step approaches use the generalized method of moments (EM-GMM) \citep{zhao2018learning} and composite marginal likelihood (EM-CML) \citep{zhao2022learning}, to avoid directly maximizing the true joint log-likelihood \eqref{M_ll}. \citet{tkachenko2016plackett} incorporated features $\boldsymbol{x}_i \in \mathbb{R}^d$ via $\gamma_i^{(h)} = \exp(\boldsymbol{x}_i^\top \boldsymbol{v}^{(h)})$ with coefficients $\boldsymbol{v}^{(h)} \in \mathbb{R}^d$ to reduce computational complexity.

\section{Applications}\label{sec:app}
\begin{figure}[t]
    \centering
    \includegraphics[width=\linewidth]{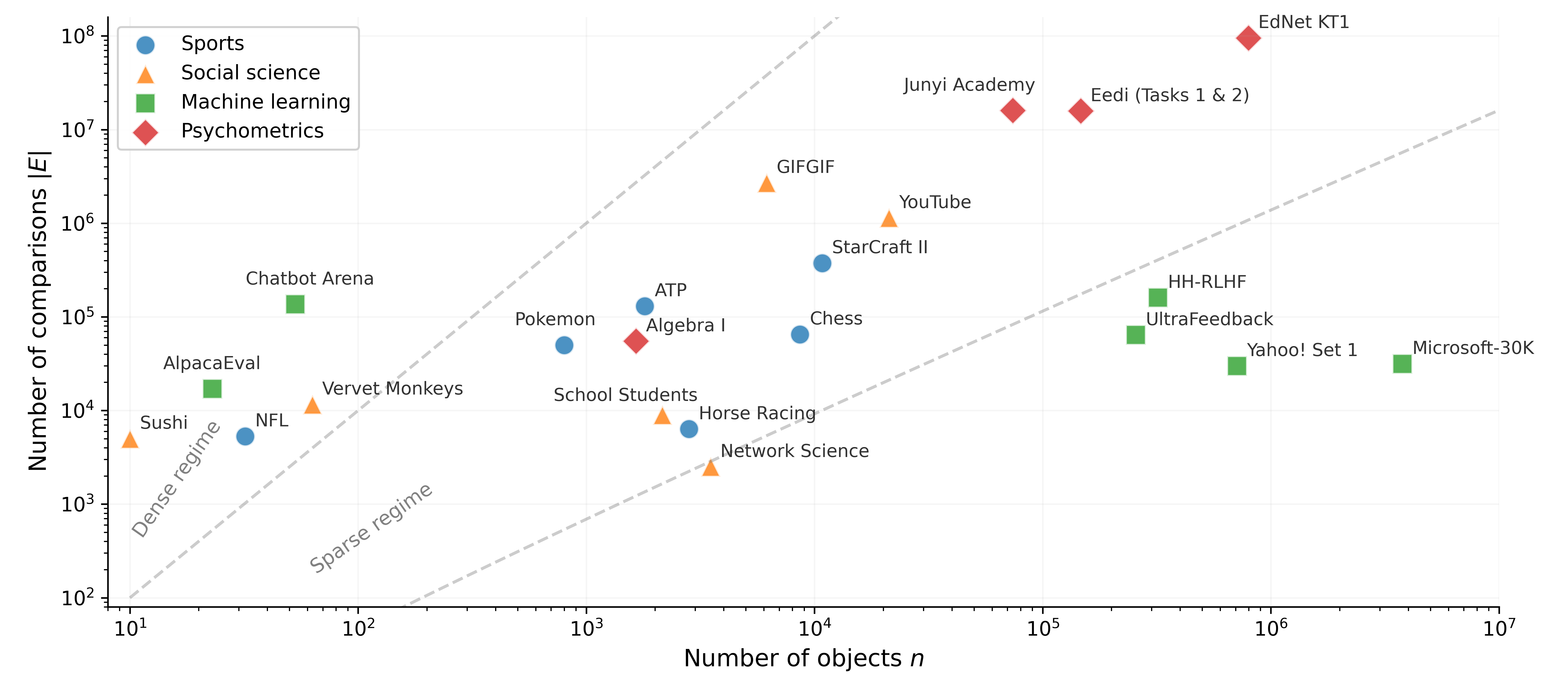}
    \caption{Overview of ranking datasets across different domains, plotted by number of objects $n$ versus number of comparisons $|E|$ on a log-log scale.}
    \label{fig:dataset_landscape}
\end{figure}

This section surveys applications of the BT model and its extensions. We focus on four areas where the BT model is extensively used: sports analytics, social choice, psychometrics, and machine learning. We examine the characteristics of the data arising in these fields and the scale of the associated problems. As illustrated in Figure~\ref{fig:dataset_landscape}, these applications span a spectrum from dense regimes where comparisons far exceed objects, say, $\mathcal{O}(n^2)$, to sparse regimes where the number of comparisons is typically of $\mathcal{O}(n\log n)$. Table~\ref{tab:datasets}  summarizes brief descriptions and key characteristics across representative datasets.

\paragraph*{Sports analytics.} 
Datasets such as American football (\citeauthor{datasetNFL}) and basketball (\citeauthor{datasetNBA}) are in the dense regime, where fixed competitor pools and frequent head-to-head matchups create highly connected graphs. In contrast, datasets such as tennis (\citeauthor{datasetATP}), {\citeauthor{datasetChess}, video games (\citeauthor{datasetStarcraftII} and \citeauthor{datasetPokemon}), and \citeauthor{datasetHorseRacing}}
fall into the sparse regime due to rapidly expanding player pools and evolving competitions. Sports outcomes are often influenced by both internal strength and the external environment; in such cases, it may be interesting, and sometimes necessary, to incorporate temporal dynamics and covariate information into the modeling.

\paragraph*{Social choice.} This domain exhibits a scale dichotomy. Datasets involving real human or animal collection, such as consumer studies (\citeauthor{datasetSUSHI}) and animal behavior (\citeauthor{datasetVervetMonkeys}), fall into the dense regime. Other datasets based on internet-collected data, including video preference (\citeauthor{datasetYouTube}), emotion perception (\citeauthor{datasetGIFGIF}), research collaboration (\citeauthor{datasetNetworkScience}), and {social network (\citeauthor{datasetSchoolStudents})}, transition to the sparse regime. As the number of objects increases, population heterogeneity becomes more prominent, and it would be helpful to incorporate such discrepancies into the modeling.

\paragraph*{Psychometrics.} Item response theory (IRT) is a widely used psychometric framework for modeling responses to test items through latent respondent traits and item characteristics. Since responses are observed between respondents and items, the underlying person-item structure is bipartite. The Rasch model, one of the simplest IRT models, shares the logistic difference structure of the BT model: the probability of a correct
response depends on the difference between respondent ability and
item difficulty \citep{rasch1960studies}. IRT datasets often exhibit substantial imbalance between the numbers of respondents and items, which can create difficulties for joint maximum likelihood estimation \citep{chen2025item}. Recently, \cite{yang2026random} proposed a debipartization approach that converts person--item responses into an item--item network to address this imbalance. Large-scale online assessment platforms provide abundant person--item response data for such analyses, such as \citeauthor{dataset_irt_kdd_algebra1,dataset_irt_neurips_edu,dataset_irt_ednet}. See \citet{dataset_irt_PSLC} for a comprehensive repository.

\paragraph*{Machine learning.} Machine learning (ML) applications are predominantly characterized by large scales, extreme sparsity, and high-dimensional feature representations. Examples include web search ranking datasets such as \citeauthor{datasetMicrosoft30K} and \citeauthor{datasetYahooSet1} for information retrieval (IR), as well as human-annotated response comparisons such as \citeauthor{datasetHHRLHF} and \citeauthor{datasetUltraFeedback} for LLM alignment. Unlike in sports analytics and the social sciences, comparisons in ML are typically conditioned on specific contexts, such as queries or prompts, decomposing the comparison graph into disjoint subgraphs. Such disconnectivity is not problematic when utilities are modeled through a shared context-dependent scoring function rather than as unrestricted object-specific parameters. This structural shift transforms the problem from parameter estimation over a connected comparison graph to function learning from disconnected local comparisons. Notably, \citeauthor{datasetChatbotArena} and \citeauthor{datasetAlpacaEval} represent a different setting, where comparisons are aggregated across prompts to evaluate a common set of LLMs, thereby inducing a substantially more connected comparison graph.

\spacingset{1.2}

\begin{table}[htb!]
\centering
\begin{threeparttable}
\footnotesize
\begin{tabularx}{\linewidth}{l c r r X}
\toprule
{\bf Dataset} & $|e|$ & $|E|$ & $n$ & \textbf{Description} \\
\midrule
\textit{Sports analytics} & & & & \\
\cmidrule{1-1}
\citeauthor{datasetATP} & 2 & 130K & 1.8K & Association of Tennis Professionals matches \\
\citeauthor{datasetClubFootball} & 2 & 230K & 1.2K & Club Football matches across 42 leagues \\
\citeauthor{datasetNBA} & 2 & 72.6K & 30 & National Basketball Association matches \\
\citeauthor{datasetNFL} & 2 & 5.3K & 32 & National Football League matches \\
\citeauthor{datasetHorseRacing} & 10--14 & 6.3K & 2.8K & Hong Kong horse racing dataset \\
\citeauthor{datasetPokemon} & 2 & 50K & 800 & Pokemon battles dataset \\
\citeauthor{datasetStarcraftII} & 2 & 374K & 10.8K & Starcraft II matches for professional players \\
\citeauthor{datasetChess} & 2 & 65K & 8.6K & Chess matches for top players \\
\midrule
\textit{Social science} & & & & \\
\cmidrule{1-1}
\citeauthor{datasetSUSHI} & 10 & 5K & 10 & A questionnaire survey of preference in sushi \\
\citeauthor{datasetYouTube} & 2 & 1.13M & 21.2K & YouTube Comedy Slam preference data \\
\citeauthor{datasetGIFGIF} & 2 & 2.7M & 6.2K & Emotion labels assigned to individual images \\
\citeauthor{datasetNetworkScience} & $2$--$14$ & 2.5K & 3.5K & Collaborations author-order data \\
\citeauthor{datasetVervetMonkeys} & 2 & 11.6K & 63 & Agonistic interactions among wild Vervet Monkeys \\
\citeauthor{datasetSchoolStudents} & 2 & 9.0K & 2.2K  & US high and middle school friendship network \\
\midrule

\textit{Psychometrics} & & & & \\
\cmidrule{1-1}
\citeauthor{dataset_irt_kdd_algebra1} & 2 & 813K & 1.65K & Student performance on Algebra I  \\
\citeauthor{dataset_irt_neurips_edu} & 2 &  15.8M & 146K  & The NeurIPS 2020 education challenge  \\
\citeauthor{dataset_irt_ednet} & 2 & 95.2M & 797K & A collection of students' question-solving logs \\
\citeauthor{dataset_irt_junyi} & 2 & 16M & 74K &  Students' exercise attempts on a learning platform.\\
\midrule

\textit{Machine learning} & & & & \\
\cmidrule{1-1}
\citeauthor{datasetMicrosoft30K} & Top-$k$ & 31.5K & 3.77M & Microsoft 30K datasets on learning to rank \\
\citeauthor{datasetYahooSet1} & Top-$k$ & 29.9K & 710K & Yahoo! Challenge Set 1 on learning to rank \\
\citeauthor{datasetHHRLHF} & 2 & 160K & 320K & Helpful and Harmless RLHF dataset \\
\citeauthor{datasetUltraFeedback} & 4 & 64K & 256K & A fine-grained and diverse preference dataset \\
\citeauthor{datasetChatbotArena} & 2 & 136K & 53 & Crowdsourced human preferences for LLMs \\
\citeauthor{datasetAlpacaEval} & 2 & 17K & 23 & An LLM-based automatic evaluation dataset \\
\bottomrule
\end{tabularx}
\begin{tablenotes}
\footnotesize
\item $|e|$: edge size, $|E|$: volume of edges, $n$: the number of objects.
\end{tablenotes}

\caption{Overview of ranking datasets}
\label{tab:datasets}
\end{threeparttable}
\end{table}

\spacingset{1.0}

\section{Conclusion}\label{sec:con}

This paper reviews several recent developments related to the BT model. Despite significant progress in making the BT model amenable to modern comparison data structures, many interesting questions remain open. Here, we list a few that are relevant to the themes of this article.

A fundamental open problem in the statistical theory of the BT model is the development of a unified asymptotic theory (e.g., uniform consistency of the MLE) that applies to heterogeneous comparison graphs. Although a number of special cases have been studied, a systematic theory capable of handling general graph structures remains open. Establishing such a framework would substantially extend the theoretical guarantees of the model and enhance its applicability to complex, real-world networks.

Furthermore, the methodology for conducting statistical inference in covariate-assisted BT models, such as the \ref{plusdc} model, is largely undeveloped. Developing hypothesis tests and confidence intervals for covariate effects is essential for obtaining reliable rankings based on individuals' merits. Such results would also allow practitioners to quantify the impact of external factors, offering actionable insights in fields such as sports analytics for performance improvement.

For mixture BT models, although their ability to capture population heterogeneity is a significant advantage, further research is needed to develop computationally efficient algorithms with proven convergence guarantees. Complementary theoretical work on the statistical properties of these estimators is also worth pursuing.

Finally, the use of {BT-type models in LLM alignment and evaluation} opens promising directions for research. A natural next step is to systematically investigate alternative pairwise comparison models and assess whether their greater flexibility can improve training efficiency, model performance, and the representation of human preferences.

\clearpage

\appendix
\setcounter{section}{0}
\setcounter{equation}{0}
\setcounter{table}{0}
\setcounter{figure}{0}
\renewcommand{\thesection}{S.\arabic{section}}
\renewcommand{\theequation}{S.\arabic{equation}}
\renewcommand{\thetable}{S.\arabic{table}}
\renewcommand{\thefigure}{S.\arabic{figure}}

\begin{center}

{\large\bf Supplement to ``Recent advances in the Bradley--Terry model: theory, algorithms, and applications''}

\end{center}

\section{Summary of notations}

\begin{table}[htbp]
\centering
\small
\caption{Main notation used in the review.}
\label{tab:notation}
\begin{tabularx}{\textwidth}{@{}l X@{}}
\toprule
Symbol & Meaning \\
\midrule
$n$, $[n]$ & Number and index set of objects \\
$E$, $\mathcal H_n(E)$ & Comparison multiset and associated graph or hypergraph \\
$e$, $|e|$, $Y_e$ & Edge/hyperedge, its size, and its observed outcome \\
$\bm u$, $\bm\gamma$ & Utility and strength vectors, with $\gamma_i=\exp(u_i)$ \\
$\bm u^*$, $\widehat{\bm u}$ & True utility vector and its estimator \\
$\sigma(\cdot)$ & Logistic function \\
$\bm x_{e,i}$, $\bm v$ & Comparison-specific covariates and their coefficient vector \\
$r$ & Shared reward function in preference learning \\
$p_n$, $L_n$ & ER edge probability and comparisons per observed pair \\
$d_i$, $d_{\min}$, $d_{\max}$ & degree of the $i$th item, the minimum and maximum degree among the items\\
$w_{ij}$, $n_{ij}$ & Wins of $i$ over $j$ and total comparisons between them \\
$H$, $\omega_h$ & Number of mixture components and component weights \\ 
$\mathcal{L},\,\Lambda$& Graph Laplacian normalized by $n$, and degree-normalized graph Laplacian, respectively.  \\
\bottomrule
\end{tabularx}
\end{table}

\section{Model relationships}
\label{app:algorithms}

\begin{figure}[!htb]
    
    \centering
    \includegraphics[width=\textwidth]{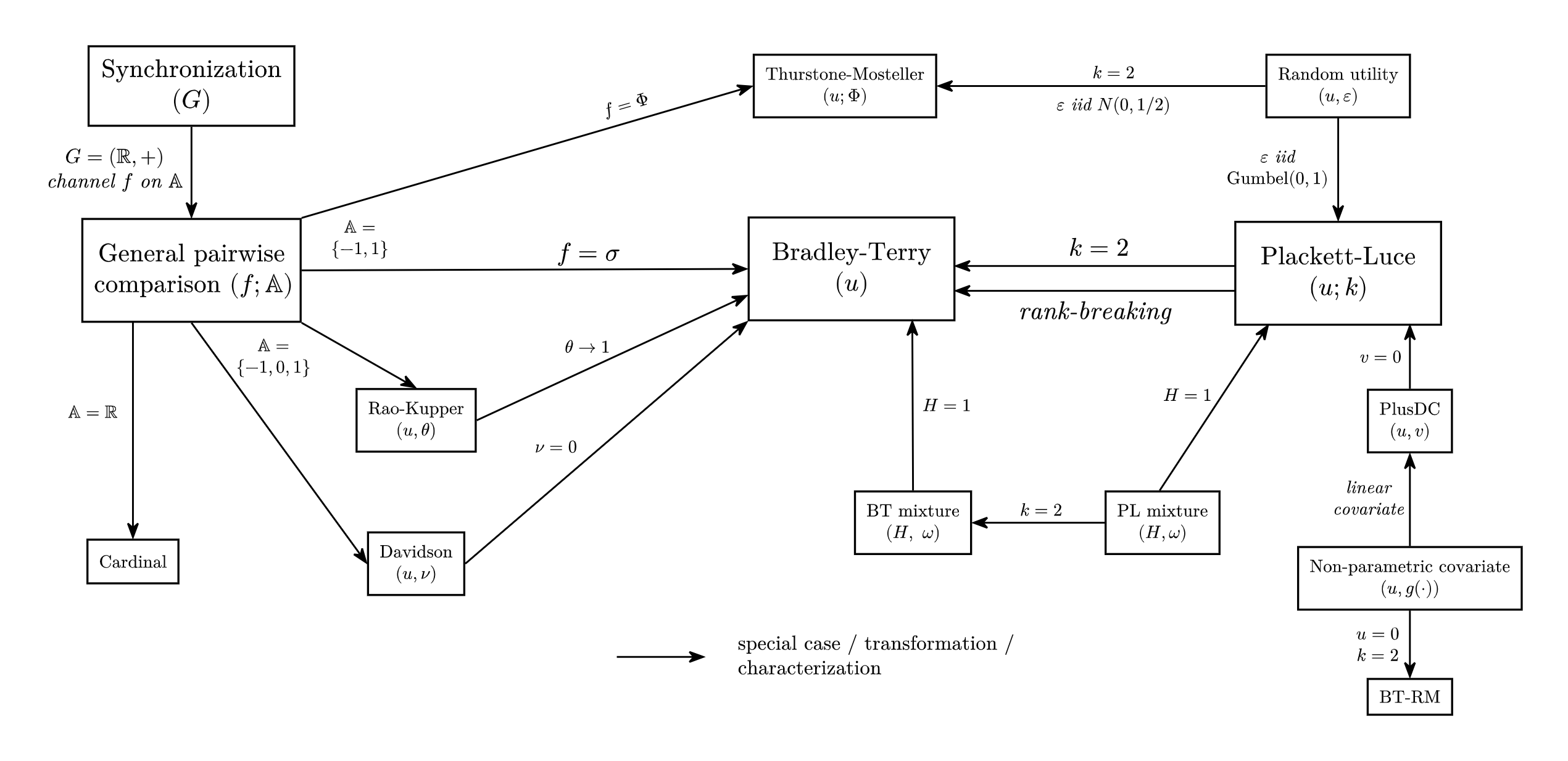}
    \caption{Relationships among some ranking models
    discussed in the main text.}\label{fig:model-relations}
\end{figure}

Figure~\ref{fig:model-relations} summarizes the relationships among the model classes discussed in the main text.

\section{Comparison of algorithms for the BT model}
\label{app:bt-optimization}

\begin{figure}[!ht]
    \centering
    \includegraphics[
        width=\textwidth,
        keepaspectratio
    ]{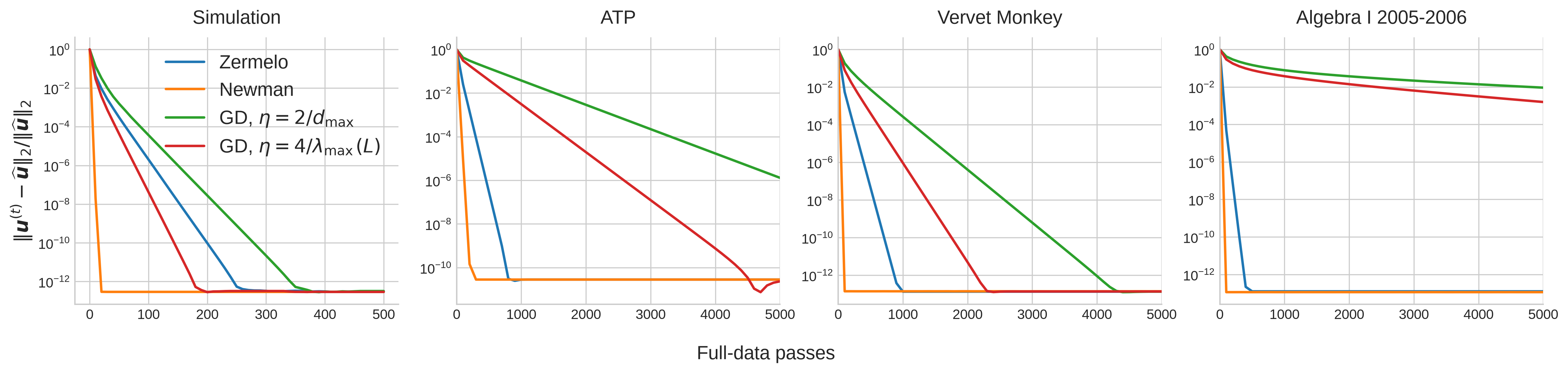}
    \caption{Relative optimization error against full-data passes. The simulation curves are averages over 100 independent replications.}
    \label{fig:bt-optimization}
\end{figure}

We compare the convergence of Zermelo's MM iteration \eqref{e3}, Newman's iteration \eqref{e4}, and gradient descent \eqref{gd} in one simulated and three real-data settings (see Figure~\ref{fig:bt-optimization}). In the simulation, $n=1000$ and the comparison graph is generated from an ER model $G(n,p_n)$ with $p_n=0.1(\log n)^3/n$. The utilities are sampled independently from $\mathrm{Unif}[-1,1]$ and centered, after which the outcomes are generated according to the BT model. The corresponding results are over 100 independent replications.

\begin{figure}[!htbp]
    \centering
    \includegraphics[
        width=\textwidth,
        keepaspectratio
    ]{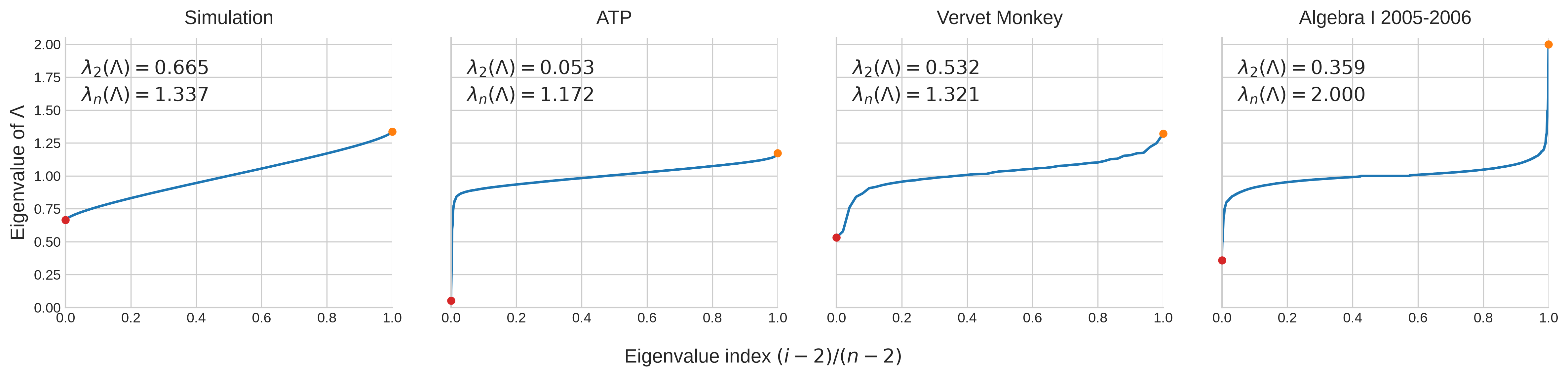}
    \caption{Ordered non-zero eigenvalues of $\Lambda$. The red and orange dots denote the spectral gap $\lambda_2(\Lambda)$ and spectral radius $\lambda_n(\Lambda)$, respectively.}
    \label{fig:spectra}
\end{figure}

The real datasets are \citeauthor{datasetATP}, \citeauthor{datasetVervetMonkeys}, and \citeauthor{dataset_irt_kdd_algebra1}, see Figure~\ref{fig:graph_vis} for their visualization. For each dataset, we prune low-degree items while preserving strong connectivity of the directed win graph, yielding $d_{\min}>30$. Let $A$ be the symmetric weighted adjacency matrix, where $A_{ij}$ is the number of comparisons between items $i$ and $j$, and let
\[
D=\operatorname{diag}(A\boldsymbol{1}),\qquad L=D-A,\qquad \Lambda =D^{-1/2}LD^{-1/2}.
\]
Table~\ref{tab:bt-graph-summary} and Figure~\ref{fig:spectra} summarize the resulting degree and spectral characteristics.

\begin{table}[!htbp]
\centering
\caption{Spectral and degree characteristics of the processed comparison graphs.}
\label{tab:bt-graph-summary}
\begin{tabular}{lrrrr}
\toprule
Setting & $n$ & $|E|$ & $\lambda_2(\Lambda)$  & $d_{\max}/\lambda_2(L)$ \\
\midrule
Simulation
& 1000 & 16,481 & 0.665 & 3.42 \\
ATP
& 1103 & 125,137 & 0.053  & 157.35 \\
Vervet Monkey
& 52 & 10,148 & 0.532 & 18.26 \\
Algebra I 2005--2006
& 886 & 676,699 & 0.359  & 326.01 \\
\bottomrule
\end{tabular}
\end{table}

Figure~\ref{fig:spectra} reveals substantial differences among the four comparison graphs. The simulated ER graph is homogeneous and has a large normalized spectral gap. ATP has a substantially ATP has a substantially smaler normalized spectral gap $\lambda_2(\Lambda)$  ($0.053$, compared with $0.359$--$0.665$ for the other graphs), suggesting
the presence of a low-conductance cut and hence relatively weak connectivity between parts of the comparison graph. Vervet Monkey is well connected and has an intermediate degree--spectral ratio. Algebra is bipartite (as $\lambda_n(\mathcal L)=2$) and is reasonably well connected, but exhibits extreme degree heterogeneity. These diagnostics suggest that ATP and Algebra are less favorable for methods using a single global step size.

All methods are initialized at $\boldsymbol{u}^{(0)}=\boldsymbol{0}$. Equations~\eqref{e3} and \eqref{e4} are implemented using asynchronous updates. Gradient descent is evaluated using both $\eta=2/d_{\max}$ and the sharper spectral step $\eta=4/\lambda_{\max}(L)$ \citep{vojnovic2023accelerated}. One full-data pass is one complete coordinate cycle for \eqref{e3} and \eqref{e4}, or one full-gradient evaluation for \eqref{gd}. We report $\|\boldsymbol{u}^{(t)}-\widehat{\boldsymbol{u}}\|_2/\|\widehat{\boldsymbol{u}}\|_2$, where $\widehat{\boldsymbol{u}}$ is the centered MLE computed to a numerical tolerance of $10^{-12}$.

\eqref{e4} converges in the fewest full-data passes in all four settings. \eqref{e3} is also substantially faster than gradient descent on the three real datasets, whereas gradient descent is comparatively competitive on the ER graphs. Using the spectral step improves gradient descent over the degree-based step, but the gap remains pronounced for \citeauthor{datasetATP} and \citeauthor{dataset_irt_kdd_algebra1}. Empirically, this pattern aligns with the graph diagnostics: gradient descent performs better when the normalized spectral gap is large and $d_{\max}/\lambda_2(L)$ is small, and worsens in the presence of connectivity bottlenecks or severe degree heterogeneity.

\subsection{Summary of Algorithms}

Table~\ref{tab:algorithms} summarizes the statistical targets, applicable models, and computational costs of the methods discussed in Section~\ref{sec:alg}.

\begin{table}[!htbp]
\centering

\scriptsize
\setlength{\tabcolsep}{3.5pt}
\renewcommand{\arraystretch}{1.05}

\begin{threeparttable}

\caption{Summary of algorithms for BT and PL models.}
\label{tab:algorithms}

\begin{tabularx}{\textwidth}{@{}
>{\RaggedRight\arraybackslash}p{0.25\textwidth}
>{\RaggedRight\arraybackslash}p{0.14\textwidth}
>{\RaggedRight\arraybackslash}p{0.25\textwidth}
>{\RaggedRight\arraybackslash}X
@{}}
\toprule
\textbf{Method}
& \textbf{Objective}
& \textbf{Models}
& \textbf{Computational cost} \\
\midrule

\multicolumn{4}{@{}l}{\textit{Likelihood and spectral methods}} \\
\cmidrule(lr){1-4}

Zermelo's/MM algorithm
\newline
\citep{hunter2004mm}&MLE& BT, PL, and selected generalized extensions & $O(|E| +n)$ per  MM iteration \\
\addlinespace[3pt]
Newman's fixed-point iteration (BT) and PL extension
\newline
\citep{NewmanJMLR2023,han2026newman,yeung2025efficient} & MLE & BT and PL &
$O(|E| +n)$ per fixed-point iteration
\\ \addlinespace[3pt]

Gradient-based
\newline
\citep{vojnovic2023accelerated} & MLE and MAP & BT, PL, and smooth extensions & $O(|E|+n)$ per full-gradient step \\ \addlinespace[3pt]

RC
\newline
\citep{Negahban2012RankCR} & Spectral estimate & BT & $O(\min\{|E|,n^2\}+n)$ per power iteration \\ \addlinespace[3pt]

ASR
\newline
\citep{agarwal2018accelerated} & Spectral estimate & PL/MNL & $O(|E|+n)$ per message-passing iteration \\ \addlinespace[3pt]

LSR and I-LSR
\newline
\citep{maystre2015fast} & LSR: spectral estimate; I-LSR: MLE at convergence & BT and PL & LSR: $O(\min\{|E|,n^2\}+n)$ per power iteration; I-LSR: $O(|E|+n+S_{\mathrm{stat}})$ per outer iteration
\\

\midrule
\multicolumn{4}{@{}l}{\textit{Bayesian methods}} \\
\cmidrule(lr){1-4}

EM-MAP
\newline
\citep{caron2012efficient} & MAP & BT and PL with Gamma priors & $O(|E|+n)$ per EM iteration
\\ \addlinespace[3pt]

Power expectation propagation
\newline
\citep{guiver2009bayesian} & Posterior approximation & PL with partial rankings & $O(|E|+n)$ per factor sweep \\ \addlinespace[3pt]

Gibbs sampling \citep{caron2012efficient}
and HMC \citep{phelan2018hierarchical}
& Posterior samples & Gibbs: BT and PL; HMC: hierarchical BT & Gibbs: $O(|E|+n)$ per sweep; HMC: $O \left(K_{\mathrm{HMC}}(|E|+n)\right)$ per HMC transition
\\

\midrule
\multicolumn{4}{@{}l}{\textit{Mixture-model methods}} \\
\cmidrule(lr){1-4}

EM-MM and EM-LSR
\newline
\citep{gormley2008exploring,nguyen2023efficient}& Mixture MLE& PL mixtures& E-step: $O(H|E|)$; M-step: $H$ weighted componentwise MM or I-LSR fits\\ \addlinespace[3pt]

EM-GMM and EM-CML
\newline
\citep{zhao2018learning,zhao2022learning,nguyen2023efficient}& Moment or composite-likelihood estimate& PL and selected random-utility mixtures& $O(H(|E|+n))$ per data pass; inner optimization omitted\\

\bottomrule
\end{tabularx}

\begin{tablenotes}[flushleft]
\scriptsize \item[] Note: We assume edge sizes are bounded and thus omitted. For spectral methods, the stated cost is per power iteration after construction of the transition matrix. $S_{\mathrm{stat}}$ denotes the cost of one stationary-distribution solve, and $K_{\mathrm{HMC}}$ denotes the number of gradient evaluations per HMC transition.  Preprocessing, evaluation costs, and implementation-dependent constants are omitted.
\end{tablenotes}

\end{threeparttable}
\end{table}

\clearpage

\bibliography{bibliography.bib}
\end{document}